\documentclass[]{interact}

\usepackage{algorithmic}
\usepackage{algorithm}
\usepackage{epstopdf}% To incorporate .eps illustrations using PDFLaTeX, etc.
\usepackage{subfigure}% Support for small, `sub' figures and tables
\theoremstyle{plain}% Theorem-like structures provided by amsthm.sty

\theoremstyle{definition}

\theoremstyle{remark}

\usepackage[numbers]{natbib}

\begin{document}
	
	\articletype{ARTICLE TEMPLATE}
	
	\title{Demand forecasting in supply chain: The impact of demand volatility in the presence of promotion}
		\maketitle
	\author
		\name{Mahdi~Abolghasemi\textsuperscript{a}\thanks{Corresponding author Email: mahdi.abolghasemi@newcastle.edu.au}, Richard Gerlach\textsuperscript{b}, Garth Tarr \textsuperscript{c}, Eric Beh\textsuperscript{a}}
		\affil{\textsuperscript{a}School of Mathematical and Physical Sciences, The University of Newcastle, NSW, Australia; \textsuperscript{b}The university of Sydney Business School, Sydney, NSW, Australia;\\  \textsuperscript{c}School of Mathematics and Statistics, The university of Sydney, Sydney, NSW, Australia}
%	}

	\begin{abstract}
		The demand for a particular product or service is typically associated with different uncertainties that can make them volatile and challenging to predict. Demand unpredictability is one of the managers' concerns in the supply chain that can cause large forecasting errors, issues in the upstream supply chain and impose unnecessary costs. We investigate 843 real demand time series with different values of coefficient of variations (CoV) where promotion causes volatility over the entire demand series. In such a case, forecasting demand for different CoV require different models to capture the underlying behavior of demand series and pose significant challenges due to very different and diverse demand behavior. We decompose demand into baseline and promotional demand and propose a hybrid model to forecast demand. Our results indicate that our proposed hybrid model generates robust and accurate forecast across series with different levels of volatilities. We stress the necessity of decomposition for volatile demand series. We also model demand series with a number of well known statistical and machine learning (ML) models to investigate their performance empirically. We found that ARIMA with covariate (ARIMAX) works well to forecast volatile demand series, but exponential smoothing with covariate (ETSX) has a poor performance. Support vector regression (SVR) and dynamic linear regression (DLR) models generate robust forecasts across different categories of demands with different CoV values.
	\end{abstract}
	
	% % % % % % % % % % % % % % % % % % % % % % % % % % % % % % % % % % % % % % % % %

	\section{Introduction}

%	Sales forecasts form an important class of information in supply chain management (SCM) upon which many decisions such as production planning, logistics, raw material purchasing, and inventory management at both the operational and organizational levels are based. In an effective SCM, it is important to know the characteristics of demand and find an appropriate forecasting model \cite{christopher2000agile}. However,	it is usually difficult to carry out forecasting with a desired level of precision because of the volatility and varying uncertainties involved \citep{jung2012managing}. 

%	However, demand forecasting can be challenging as there are many uncertainties associated with them \cite{syntetos2016supply}.
	
	Demand is one piece of the important information that can be shared and used in supply chain management (SCM). Demand sharing and demand forecasting are extremely helpful for supply chain managers since it provides a great source of information for planning and decision making. Demand forecasting is the basis for a lot of managerial decisions in the supply chain such as demand planning \cite{narayanandemand}, order fulfilment \cite{narayanandemand}, production planning \cite{donohue2000efficient}, and inventory control \cite{silver1998inventory}. It is usually difficult to carry out forecasting with a desired level of precision because of the volatility and varying uncertainties involved \cite{jung2012managing,syntetos2016supply}.
	
%	, budgeting \cite{jordan2019use},
	Demand volatility inherently exists due to the consumers' behavior that is constantly changing \cite{walker1984transaction}. Various variables such as promotion, weather, market trends, and season may have an impact on consumers behavior and contribute to demand volatility \cite{gilliland2010business}. Promotion, in particular, is a very common practice in the retailing industry that can make demand volatile. Promotions impact on demand dynamics have been investigated extensively in the literature (see, for example, \cite{Ali2009,Divakar2005,Nikolopoulos2015,Ramanathan2012,Ramanathan2011,Trapero2014}). One natural outcome of the promotion is demand volatility. Demand volatility and variability can occur before, after or during promotional periods \cite{blattberg1993}. 
%	 Volatility and uncertainty along with complexity and ambiguity are considered as four main factors affecting the supply chain performance \cite{chen2000quantifying,zhang2004impact}.

	Demand volatility is a challenging risk to supply chain and it is becoming a concern for managers and practitioners \cite{christopher2000agile}. Researchers and practitioners have repeatedly raised their concerns about increasing demand volatility as a threatening risk to the supply chain \cite{christopher2017supply}. Demand volatility renders demand forecasting a difficult task and poses excess costs for stock-outs, inventory, and capacity utilization \cite{christopher2011supply}. However, demand volatility has been under-considered in supply chain demand forecasting literature. Forecasting to capture the underlying behavior of the volatile demand is crucial to diminish the uncertainty in different levels of the supply chain.
	There is no single model that performs well for all different types of demand series. Often a combination of judgment and statistical model is used to forecast the volatile demand series that are impacted by promotion. For example, rule-based forecasting is one of the methods that combines experts knowledge and forecasting techniques based on 99 rules to generate forecasts that are suitable for the feature of data \cite{collopy1992rule}. We can identify the appropriate forecasting model based on time series characteristics \cite{wang2009rule,wang2006characteristic,fulcher2013highly}. 
	To the best of our knowledge, there is a lack of empirical studies that investigate the volatility of demand caused with promotion as a criterion to develop a forecasting model. 
%	The most suitable forecasting method under different conditions has been recommended by researchers in different contexts. (See, for example, \cite{graff2013models,wang2009rule,collopy1992rule,garland2014model,armstrong2001principles}). 

%	compare the relative variation between different series.
%These models are often used to forecast sales during promotional weeks, or at best over a period before and after promotion. However, the entire demand series should be considered in developing a forecasting model as the promotion may impact demand vastly.

	We use CoV to measure the volatility of demand and propose appropriate forecasting models when demand series exhibit different values of CoV. CoV by definition is the sample standard deviation divided by the sample mean. CoV is a scale-independent metric and can be used to compare the relative variation across multiple series. CoV represents the uncertainties in data and it is considered as the demand variability criteria in supply chain \cite{packowski2013lean, cachon1999managing}. We use this metric to measure the promotional variations in demand as well as the natural variations of demand. This is because promotion does not only impact demand over the promotion periods, but changes demand over a number of periods. In general, demands with large CoV values are associated with more uncertainties and are difficult to forecast \cite{huang2008demand}. These demands, if not forecasted accurately, can cause many problems in the upstream supply chain operations; and pose a shock to supply chain activities as they can take diverse forms. We aim to find the most suitable model for different values of volatilities. 

%	
% . This included convenience to use, market popularity,  structured judgement based on experts advice, statistical criteria based on what should work, relative works based on what has worked previously, and guidelines from prior research that what works in this type of situation. 

	The main contributions of this study are two-fold: methodological and practical. The methodological contribution is firstly to introduce CoV as a criterion that should be considered in developing appropriate forecasting models and, consequently, generating robust forecasts. We argue that demands with different CoV require different types of a forecasting model that suits their characteristics. Secondly, we show that a decomposition-based model works well to forecast the volatile demand. We decompose demand to the non-promotional (baseline) and promotional demand where each of them has a smaller CoV. Then, we propose a hybrid model that uses a piecewise regression model and a time series model to forecast the promotional and baseline demand, respectively. We compare this model with other well-known models in the literature and show that our model has several advantages to the other existing models in the literature in terms of accuracy, simplicity, robustness to volatility, and ability to have a prediction interval. Finally, we develop and implement a number of well-known forecasting methods to evaluate their performance when demand series exhibit volatility.

	We also make a number of practical contributions to the field. First, considering the CoV of demand series, managers will be able to use an appropriate model to capture the volatility of demand automatically with a high level of accuracy as opposed to judgmentally overriding the output of statistical models. Second, by using our proposed models, experts will save a substantial amount of time as it mitigates the need for labor-intensive judgmental forecasts. Third, we develop forecasting models that only use price as the independent variable. These models are simple, intuitive, and easy to use in practice that make them desirable for practitioners \cite{armstrong2001principles}. Finally, the accuracy of the forecasts lead to more efficient source utilization of costs, reduces the inventory level, and improves the performance of the upstream supply chain substantially.

%	 This will consequently help managers in the upstream SC to make production planning and materials resource planning decisions easily. 
	The remainder of the paper is organized as follows. Section \ref{litrev} reviews the literature. Section \ref{Methodology} discusses the methodology and forecasting techniques that are used to forecast demand. Section \ref{data} describes our data set and  Section \ref{resuts} provides the results and discussion. Conclusions are drawn in section \ref{conc}.
	
	\section{Literature review} \label{litrev}
	Supply chains are associated with many uncertainties and complexities in the modern world. These uncertainties arise for many reasons such as supply chain network, the activities of partners, customers behavior, competitors behavior, emerging technologies, and new product development all of which contribute to a volatile supply chain \cite{christopher2011supply,serdarasan2013review}. A volatile market causes a volatile demand. In order to avoid the negative consequences of volatile demand, it is important to consider and predict the uncertainties.

	Demand forecasting is one of the factors that can contribute to the volatility of demand \cite{gilliland2010business}. Demand forecasting is not an easy task and many companies and forecasters fail to do a scientific forecast \cite{armstrong2017demand}. The biggest problem with demand forecasting is the uncertainty in demand that renders demand forecasting a challenging problem \cite{syntetos2016supply}. Demand accuracy is a critical factor in determining the quality of decision making \cite{jordan2019use}. Inaccurate forecasts may cause unnecessary costs in procurement and transportation, manpower, service level, and inventory \cite{chopra2004supply,torkul2016real}. Although demand volatility can be reduced and controlled, it is inevitable \cite{volatilitythesis}. Thus, it is crucial to have a proper strategy to control volatility.
	
%	Volatile demands are often uncertain and more difficult to forecast, but demand volatility does not necessarily render forecasting difficult. For instance, a seasonal demand might have high volatility but it may not be difficult to forecast. 

%	In this thesis, we take the planning approach rather than controlling. Controlling the volatility can become obsolete in an uncertain environment where companies compete in a dynamic market \cite{hope2003beyond}. Therefore, we aim to model and forecast the volatile demand so that supply chain managers can plan for it in advance. 
	Researchers and practitioners have proposed and developed techniques and approaches to deal with demand volatility. For example, one way to avoid the negative impact of demand volatility is to increase the inventory level. This will help to counter the demand volatility but imposes a lot of costs to the companies \cite{chopra2004supply}. Another strategy to control demand volatility is to increase the capacity, but this is not an attractive proposition as it costs a lot for the supply chain. These methods can be useful to control the volatility of demand. However, they may not be cost-effective.

%	 During the last couple of decades, various quantitative and qualitative forecasting models are developed and applied to forecast demand \cite{Aye2015, Pai2005,fildes2008forecasting,hyndman2014forecasting,syntetos2016supply,lawrence2006judgmental}. However, 

%	Consequently, 
	Demand forecasting is the prerequisite for strategies that aim to control the volatility of demand \cite{hope2003beyond}. It is the first step towards dealing with uncertainty and volatility in the supply chain. Over the past few decades, many different models have been used for retail sales forecasting \cite{Aye2015, Pai2005,fildes2008forecasting,hyndman2014forecasting,syntetos2016supply,lawrence2006judgmental}. There is no unique solution that can address all types of forecasting problems and performs better than all other forecasting models in all possible situations and under all possible conditions. However, some models might outperform others under particular conditions. For instance, \cite{alon2001forecasting} compared the forecasts generated by an artificial neural network (ANN), triple exponential smoothing, ARIMA, and multiple aggregation methods. They concluded that ARIMA and triple exponential smoothing outperform the other models when the macroeconomic conditions are stable, whereas ANN and multiple aggregations might work better in volatile markets. In other studies, \cite{meade2000evidence} and \cite{shah1997model} considered CoV as the statistics in analyzing different time series models. However, they did not investigate different models performance with respect to their CoV as an important feature of demand series in SC context that can impact the forecasting performance. These methods underestimate or neglect volatility as a factor that can impact the performance of the forecasting models. We shall construct or choose an appropriate model based on the characteristics of the demand. 
	 %We review a range of forecasting model and assess them on our case study.
	 
%	Extrapolative methods are popular forecasting methods assume that the past pattern of a series is representative of its future. Therefore, we can estimate the future of the series based on its relationship with other variables. These models are one of the very first developed forecasting models. A good review of this models is presented by \cite{fildes1992evaluation}. We use ARIMA and exponential smoothing as two of popular and powerful extrapolative models \cite{hyndman2014forecasting, hyndman2007automatic}.  They have been applied to various forecasting problems including forecasting electricity load \cite{caprio1983short}, replenishment orders \cite{sillanpaa2018forecasting}, and M-competitions \cite{makridakis2018}. 

%	  Exponential smoothing models work better than ARIMA when data are non-normal \cite{Hyndman2008}.
%	\citet{Hyndman2001} developed state space models to derive prediction interval for some of exponential smoothing models. 
 	Extrapolative methods are popular forecasting methods that assume the past pattern of a series is representative of its future. Therefore, we can estimate the future of the series based on its relationship with other variables. They have been applied to various forecasting problems \cite{caprio1983short,sillanpaa2018forecasting}, and can be used as benchmark models \cite{makridakis2018}. We use ARIMA and exponential smoothing as two of popular and powerful extrapolative models \cite{Billah2006,hyndman2014forecasting, hyndman2007automatic}. These methods rely on auto-correlations with past data and base their forecasts on extrapolations from past patterns. Hence, they work well only when the future is similar to the past; indeed, the pattern changes severely compromise their efficacy. Conventional time series models lack the ability to adequately capture random variation in series and ignore the impact of influencing variables on demand \cite{huang2008demand}. Therefore, a univariate ARIMA or exponential smoothing model might fail to forecast well if demand time series is subject to volatility. To overcome this problem, one can add influential variables to time series and construct ARIMAX and ETSX models. These models consider both autocorrelations and influential factors and are promising approaches to the time series literature. \cite{athanasopoulos2008modelling}. We implement these models to evaluate their performance for the volatile demand series.

	Econometric forecasting models have been developed to address the pitfalls of conventional extrapolative forecasting models.  Causal and regression models are a very popular form of econometric models. Various forms of causal models have been developed for retail sales forecasting \cite{Science2016a,Ali2009,Christen2016,Christen2016}. More recently, other sophisticated regression models such as SCAN*PRO \cite{van2002promotions} and CHAN4CAST \cite{Divakar2005} have been proposed to capture promotion impact. These models are static and designed to work at the brand level. The downside of these models is that they are expensive to use as they need many inputs variables. Moreover, it is not clear if these models have a robust performance to forecast volatile demand across the entire demand series and not only during the promotion as a special event. The retail market is a dynamic environment in which the demand time series is changing continuously due to many different factors of major and minor influence. Therefore, a dynamic approach may fit and forecast the volatile demand better.

%	Demand variability and demand forecastability are two of important factors to classify demand \cite{packowski2013lean}. 
	Demand series with high volatility may take different values over the horizon. One can use different models to forecast similar parts of demand series more accurately, and then combine them to forecast the entire demand series \cite{Aye2015,Pai2005}. Hybrid models that use combination approach have been successfully implemented in retail sales forecasting \cite{Aye2015,Pai2005,guo2017double}. An empirical study on the retail industry has shown that a hybrid ARIMA and ANN model improves the accuracy of demand forecasts \cite{Aburto2007}. In other research, a hybrid SVR and ARIMA model outperformed the standard statistical models \cite{Pai2005}. We develop a hybrid model that first decomposes demand into main parts and forecast them separately, and then combines them to forecast the demand.
	
%	Point forecasts do not provide any information about the associated uncertainty in forecasts. Moreover, there is no evidence that how they perform in forecasting volatile demand series. 
	
%	Hybrid models are another type of forecasting models that are developed to combine different models such that they can take advantage of their strength. .. 
%	We can use different methods for combining forecasts \cite{clemen1989combining}. The combination can be used at the beginning of modelling by decomposing series into their components and constructing a hybrid model. Alternatively, the forecaster may use individual models to forecast series and take average from the generated results.  

%	Hence, practitioners cannot assess the forecasting methods based on their uncertainty and choose the most appropriate one according to the acceptable associated risk. Managers are interested in an interval as it provides a range for forecasts and can be helpful for decision making in operations management and upstream supply chain.
	
%	 We develop a decomposition based model called HR-ARIMA model that has a prediction interval. The performance of this model is evaluated for series with different level of volatility. \\	
	
%	ML algorithms are another emerging technique that has experienced rapidly growing interest in retail sales forecasting and are serious competitors with classical statistical models. They can learn from data, recognise patterns and generalise outcomes to make projections about the future. 

	During the last decades, ML algorithms have gained a lot of attention in demand forecasting \cite{priore2018applying,ghiassi2009dynamic}. However, they have not been fully explored in SCM context and require more attention \cite{min2010artificial}. ML algorithms are computationally more expensive but provide a range of different and flexible models for forecasting demand. This may help them to forecast the volatile demand series easier. Much debate surrounds the relative performance of statistical and ML methods, and it is difficult to draw general conclusions about their efficacy \cite{Aye2015, makridakis2018}. Each class of models might outperform others under certain conditions \cite{ahmed2010empirical}. We utilize ANN and SVR as two of the most common and successful ML techniques for retail sales forecasting \cite{makridakis2018, min2010artificial}. The performance of these models has not been fully explored when they face volatile demand time series. We implement these models on our data set and compare the results obtained using these techniques with those obtained using a number of statistical models.

	We empirically investigate a large data-set of demand time series that have demands with different levels of volatilities to develop a forecasting model. On one side, we aim to improve the forecasting practice for volatile demand series and on the other side give insight to managers in similar firms to cope with demand volatility in the supply chain. Furthermore, the empirical results drain in this paper can be used to develop theories around demand volatility in supply chain forecasting model. 
	
%	There are many different models proposed with different ML algorithms but limited evidence is available about their performance in demand forecasting with relative to statistical models \citep{makridakis2018}. We have implemented two common ML algorithms to compare their performance with other statistical models.

%	There is no unique solution for forecasting problems that can address all types of forecasting problems and performs better than all other forecasting models in all possible situations and under all possible conditions. There have been a number of studies exploring what methods work better than others for certain types of time series with certain characteristics \cite{kang2017visualising, meade2000evidence, wang2009rule}. For instance, the forecasts generated by an ANN, triple exponential smoothing, ARIMA and the multiple aggregation methods \cite{alon2001forecasting} have been compared. Researchers found that ARIMA and triple exponential smoothing outperform the other models when the macroeconomic conditions are stable, whereas ANN and multiple aggregations might work better in volatile markets. Consequently, we shall construct or choose an appropriate model based on the behavior demand. Series with different levels of volatility requires different models. We develop nine forecasting models and evaluate their performance using demand series with different levels of volatility. 

\section{Methodology}\label{Methodology}

	%As shown in Section \ref{data} sales behavior during promotions is very different from that for non-promotional periods. 
	%Conventional time series models are unable to provide accurate forecasts during promotional periods where sales are dramatically greater than non-promotional sales. One difficulty with ARIMA models is that they only use historical data and autocorrelations among the time series observations to forecast, ignoring the importance of some influencing factors which might impact sales. We shall find the most influential factors during promotional periods and embed them into our models. 
	We aim to understand how promotion and demand volatility changes the efficacy of forecasting models, and how it can be modeled in the supply chain. We underpin our arguments with evidence from hundreds of demand time series of a major fast moving consumers good (FMCG) company. We strive to gather as many data as possible to make sure the results are valid, robust and generalisable to the other similar firms.

	Many different variables can affect the dynamics of demand and it is difficult to distinguish their relative impact on demand. It may be expensive and sometimes not practical to gather all information that play role in demand. Upon availability of all information, we can build multivariate models to forecast volatile demand. Although this is a natural approach to forecast volatile demand, it is not possible to use this method as data may not be available due to different reasons. As such, it is of particular importance to choose the most influential variables that are available and are believed to drive the outcome of demand \cite{Ali2009}. In the current study, we do not have access to tidy data of all variables that may contribute to the volatility of demand. We only have access to the price of all products. The average of Pearson's correlation between price and demand is -0.83 indicating a very strong relationship which can potentially explain a high portion of demand variation. Thus, we rely on price as an influential factor to construct and evaluate different techniques for demand modeling.

	We develop a hybrid regression time series model, called HR-ARIMA. We also build different models from existing methods in the literature including ARIMAX, ETSX, DLR, ANN, and SVR. ARIMA, ETS, and \textit{`Theta'} models are used as benchmarks.  \textit{`Theta'} is one of the hybrid models based on the decomposition of time series that draw a lot of attention after it had been successfully implemented on M3-competition data \cite{makridakis2000m3}. This model decomposes the data into two \textit{`Theta'} lines. The first \textit{`Theta'} line removes the curvature of data and the second line doubles the curvature of data. The mean of these two models generates the final forecast \cite{assimakopoulos2000theta}. We discuss these models in detail in the following subsections.

\subsection{Hybrid Model} \label{heumodels}

	The hybrid model was formulated, considering the fact that the behavior of demand during promotional periods is completely different from that during non-promotional periods and they require different models as their nature is different. Therefore, we decomposed demand into two main factors that cause volatility which are baseline demands and promotional demands. One can decompose promotional demand more in detail; however, in this paper, we are interested only in the size of promotional demand and not the impact of different promotional factors on demand.
	
	Understanding and estimating baseline demand is fundamental of many analysis in marketing and the basis for estimating the promotion behavior. We used an ARIMA model to estimate baseline demand. The models are implemented in R and the `\textit{forecast}' package is used to fit and estimate the parameters of the ARIMA model \cite{hyndman2014forecasting}. Baseline demand is an estimate and not an actual number. It is difficult to measure whether it is accurate or not because there is no actual number to compare it with \cite{gilliland2010business}. 
	Other simple and sophisticated methods can be used to estimate the baseline demand \cite{Science2016a, Jetta2011}. However, implementing them will not significantly influence forecasting accuracy as demand in the absence of promotions is fairly stable and small compared to promotional demand. They are neither the forecasting concern of managers in our investigated supply chain nor the focus of this paper.

%As another attempt, we used an order three weighted moving average of past non-promotional sales to estimate the baseline sales in the absence of promotions. The generated results are very similar to the ETS model. For simplicity, as it is desirable for practitioners, we can use a simple moving average model. 

	After estimating baseline, demand uplifts are found simply by subtracting baselines demand from total demand at each period. Surprisingly, we found that there is a stronger correlation between price and demand uplift only because of promotions than between total demand and price. Pearson's correlation between demand uplift only due to promotions and price is -0.89, indicating that a linear regression model is capable of forecasting demand uplifts very well.
%This is because products have a baseline sales rate, regardless of being on promotion or not. The quantity of baseline sales is involved in the price-sale relationship when we are computing the correlation coefficient; however, it should be excluded as the baseline sales do not occur specifically because of the product's price. \\
	We developed a piecewise regression model to capture promotional uplifts. Piecewise regression is a type of polynomial regression where a number of regression models join together at knots. Knots are the points where the model parameter changes. Since promotions' prices are fixed, we set the knots at different promotional prices.

In Algorithm \ref{Heuristic Algorithm}, a stepwise procedure is proposed for the construction of such a hybrid model: \\

\begin{algorithm}
	\caption{Hybrid model algorithm}
	\label{Heuristic Algorithm}
	\begin{algorithmic}[1]
		\FOR{$t=0$ to $N$}
		\STATE Decompose demand into main components that have lower volatility. This includes baseline demand and uplift in demand only due to the promotion.\\
		\STATE Estimate the baseline demand for promotional periods.\\
		\STATE Subtract baseline demand from total demand to find the uplift size due to promotion.\\
		\STATE Set the promotional prices for each of the series to construct a piecewise regression model.\\
		\STATE Construct a piecewise regression model for demand uplifts for different promotions.
		\ENDFOR
		\STATE Forecast each of decomposed parts separately.\\
		\STATE Sum them up where appropriate. 
		%		\RETURN $S_t$, for $t=1,\dots,N$.
	\end{algorithmic}
\end{algorithm}

	 We not only separated demand during promotions from demand during non-promotional periods but also decomposed the demand during promotional periods into baseline demand and demand occurring only due to the promotion. Then, we identified the different range of promotional prices for each product and fitted a regression model for each range. Each piece of the hybrid model is shown in Equation \ref{hybrid}
 \begin{align}
 &s_t = \log(y_t)  +  \lambda_k( \alpha  + \beta \log(r_{t})) \label{hybrid} ~~~~~
 \end{align}
 
 where $\lambda_k$ takes one if promotion type $k$ is offered, and zero otherwise. $s_t$ and $r_t$ denote total demand and price, respectively.\\
 
 	This hybrid model has two components, an ARIMA, and a piecewise regression. The first component, $y_t$, is a time series model and could be any other model capable of forecasting baseline demands.  The second component is a piecewise regression model which is used for forecasting demand uplifts for different prices \cite{van2002promotions}. 
 
 %\subsubsection{Prediction Interval}
	 One advantage of this hybrid model is that we can construct a prediction interval for it. Prediction intervals will provide managers with insight into the most appropriate choice of forecasting methods when the degree of uncertainty is taken into account. This is very helpful for them to plan for promotions and control the level of inventory when they are dealing with large variations and tremendous uncertainties during promotions. Prediction intervals consist of an interval with an upper and lower limit between which a forecast is expected to lie with a certain probability. The endpoints of prediction intervals are found using the formula $y_t$=  $\hat{y_t}$ $\pm$ $k \hat{\sigma}$, where $k$ is the multiplier that determines the percentage of the prediction interval and $\hat{\sigma}$ is the standard deviation of the forecasting error. This model assumes that residuals are normally distributed and that $\hat{\sigma}$ is the standard deviation of the forecast distribution \cite{hyndman2014forecasting}. The prediction interval for HR-ARIMA is constructed in the same way as the errors of regression and time series components are independent and follow a normal distribution. If $N(0, \sigma_{rk}^2)$ denotes the distribution of the forecasting error for the ${k^{th}}$ piece of the regression component, then the forecasting error of the piecewise regression model follows the  $N(0, \sigma_{r1}^2 + ...+ \sigma_{rk}^2)$ distribution since the error terms are independent. Suppose $N(0, \sigma_t^2)$ denotes the distribution of the forecasting error for the ARIMA component, then the forecasting error of the hybrid model follows the $N(0, \sigma_{r1}^2 + ... + \sigma_{rk}^2 + \sigma_{t}^2)$ distribution since the error terms are independent.
 
%\subsection{State Space Models}\label{SSM}
%State space models (SSM) consider time series as the output of a dynamical system with random noise \citep{kalman1960new}. It provides a flexible structure for specification and captures different types of behavior in time series.
%
%Equations (1) and (2) show the general form of innovation SSM where $y_t$ denotes the observation equation and $\boldsymbol{X_t}$ represents the state equation:
%\begin{align}
%	&y_t = \boldsymbol{f_t X_t} + \varepsilon_t \label{eq:state}\\
%	&\boldsymbol{X_t}= \boldsymbol{G_t X_{t-1} }+ \boldsymbol{w}_t\\ \nonumber
%	&\varepsilon_t \sim \mathcal{N} (0,\sigma^2)\\ \nonumber
%	& \boldsymbol{w_t} \sim \mathcal{N} (\boldsymbol{0}, \boldsymbol{W_t}) \nonumber
%\end{align}
%	Here $f_t$ and $G_t$ are the observation and state transition matrices, respectively. If $\boldsymbol{w_t}$ can be written as  $\boldsymbol{g}$$\epsilon_t$, then the errors (innovation) of the observation and state equations will be the same.
%	The parameters of the SSM for the ARIMAX, ETSX, and DLR models will be estimated in subsequent subsections. We put all these models in a state space form which provides a unified and elegant structure to model and forecast sales.\\

\subsection{ARIMAX}\label{ARIMAX}

	The ARIMAX model is a generalization of the ARIMA model that can be constructed by adding an explanatory variable to the ARIMA model.  The parameters $p$, $d$, and $q$ in an ARIMA$(p,d,q)$ model represent the autoregressive (AR) order component, the order of differencing and the order of the moving average (MA) component, respectively. Equation (\ref{arima}) demonstrates an ARMA$(p,q)$ model where $e_t$ is a white noise process with mean zero and variance $\sigma^2$:

	\begin{align}
	y_t = \phi_1y_{t-1} + ...+ \phi_py_{t-p} + e_t + \theta_1 e_{t-1} + ... + \theta_q e_{t-q}. \label{arima}
	\end{align}

	When additional information is available, it is often beneficial to add the covariate to the forecasting model. ARIMAX can be constructed by adding price $r_t$ as the covariate to the right hand side of Equation (\ref{arima}).
	\begin{align}
	z_t = \beta r_t + \phi_1z_{t-1} + ...+ \phi_pz_{t-p} + e_t + \theta_1 e_{t-1} + ... + \theta_q e_{t-q} ,
	\end{align}
	where $z_t= \Delta^d~ y_t$.
	
	The models are implemented in R and the `\textit{TSA}' package is used to fit the ARIMAX model and estimate the parameters \cite{TSA}. The models with the lowest AICc are chosen as the best fitted models.
	
%	The parameters and matrices of ARIMAX model in state space form can be represented as follows: 
%\begin{align*}
%\boldsymbol{\hat{f}_t}= \begin{bmatrix}
%1 & -\theta_1
%\end{bmatrix},~ \boldsymbol{X_t'} =
%\begin{bmatrix}
%e_t& e_{t-1}
%\end{bmatrix},~\boldsymbol{w_t'} =
%\begin{bmatrix}
%e_t & 0\\
%\end{bmatrix},~\boldsymbol{W_t} =
%\begin{bmatrix}
%\sigma^2 & 0\\
%0&0
%\end{bmatrix},
%~ \boldsymbol{\hat{G}_t} =
%\begin{bmatrix}
%0 & 0 \\
%0 & 1
%\end{bmatrix},~\varepsilon_t = 0
%\end{align*}
\subsection{ETSX}\label{ETSX}

	Exponential smoothing has frequently been applied in both research and practice since its introduction by Brown in 1959 \cite{brown1959statistical}. Exponential smoothing provides a weighted moving average in which the most recent observations are given more weight. Different types of exponential smoothing models employ different parameters, which enable them to capture irregular patterns. A statistical framework for exponential smoothing called ETS has recently been developed in the state space framework. ETS methods are classified based on their components: errors, trends, and seasonality. The trend component includes a level (l) and a slope (b), which can be combined in different ways. In general, five types of trends, three types of seasonality, and two types of errors (additive and multiplicative) exist for ETS models, yielding a total of 30 different methods. The point forecasts for single and multiple sources of error are the same, but the prediction intervals are different \cite{hyndman2008forecasting}. 
%[Insert Table 2 here]\\

The following equations describe the ETS model:

\begin{align}
&Y_t = l_{t-1} + \phi b_{t-1}+ \varepsilon_t\label{observationets}\\
&l_t= l_{t-1} + \phi b_{t-1} + \alpha \varepsilon_t\label{state1ets}\\
&b_{t}=  \phi b_{t-1} + \beta \varepsilon_t\label{state2ets}\\\nonumber
\end{align} 

	Equation (\ref{observationets}) shows the observation, while Equations (\ref{state1ets}) and (\ref{state2ets}) show the state equations of the ETS model. Here, $\varepsilon_t$ is a white noise process with mean zero and variance $\sigma^2_\varepsilon$.

	When additional information is available, it can be embedded as a regressor into the ETS model. The ETSX model can be constructed by adding a covariate to the ETS to improve the forecasting accuracy. The regressor variable is added to the observation equation. In the following model, we have considered the price $r_t$ as the covariate, and given it a time-invariant coefficient of $c$:

\begin{align}
&y_t = l_{t-1} + \phi b_{t-1}+ c r _t+ \varepsilon_t \label{etsxobs}\\
&l_t= l_{t-1} + \phi b_{t-1} + \alpha \varepsilon_t\label{estxstat1}\\
&b_{t}=  \phi b_{t-1} + \beta \varepsilon_t\label{estxstat2}\\\nonumber
\end{align}

	Equation (\ref{etsxobs}) shows the observation equation and Equations (\ref{estxstat1}) and (\ref{estxstat2}) show the state equations of our ETSX model. Again, $\varepsilon_t$ is a white noise process with mean zero and variance $\sigma^2_\varepsilon$.
%The ETSX model can be expressed in general state space form as follows:
%\begin{align}
%	&Y_t= y_t - r_t c = \boldsymbol{f_t} \boldsymbol{x}_{t-1} +\varepsilon_t~,
%\end{align}
%where,
%\begin{align*}
%\boldsymbol{f_t}= \begin{bmatrix}
%1 & \phi
%\end{bmatrix},~\boldsymbol{x}_t = 
%%\boldsymbol{X_t'} =
%\begin{bmatrix}
%l_t & b_t
%\end{bmatrix},~
%\boldsymbol{w_t'} =
%\begin{bmatrix}
% \alpha & \beta
%\end{bmatrix}\varepsilon_t,~\boldsymbol{W_t'} =
%\begin{bmatrix}
%\sigma^2 & 0\\
%0 & \sigma^2
%\end{bmatrix},~
%\boldsymbol{G_t} =
%\begin{bmatrix}
%1 & \phi \\
%0 & \phi
%\end{bmatrix}
%\end{align*}
 The model and parameters are set and estimated in R using the \textit{`smooth'} package \cite{ivan}.

%	In our investigated case, the exponential smoothing works poor during promotions and experts do not rely on the statistical output. In other words, experts in industry tend to believe that statistical models are always dependent on humans judgment and they are unable to generate accurate forecast in different conditions such as promotions. However, with a small change in exponential smoothing model and embedding the covariate, they can perform well during promotions too. This change, despite its simplicity, makes significant improvement in forecasts and saves time and cost substantially against labor intensive judgmental adjustment. 

\subsection{Dynamic Linear Regression}\label{dlr}
	DLR is considered to be a generalization of the standard linear regression models and can be expressed in a state space form. Although univariate models such as stochastic volatility models might provide quite good descriptions of the series behavior with irregularities or jumps; they will rarely be capable of predicting sudden changes without further information. DLR does not assume a regular pattern and stability of the underlying system but accommodates sudden and massive changes \cite{dlmr}.
	Equation (\ref{dlmobs}) provides an observation equation for a DLR model which is initiated from the price-sale relationship. The state equations (\ref{dlmstate1}) and (\ref{dlmstate2}) are the coefficients of a linear regression model which follows independent random walks \cite{van2002promotions}:
	
	\begin{align}
	&\log(y_t) = \alpha_t  + \beta_t \log(r_{t}) + \varepsilon_t\label{dlmobs}\\
	&\alpha_t= \alpha_{t-1} + \nu_t\label{dlmstate1}\\
	&\beta_t= \beta_{t-1} + \omega_t\label{dlmstate2}\\ \nonumber
	\end{align}  
	In above equations, $r_t$ denotes the price at at time $t$, $y_t$ represents the demand at time $t$,  and $\varepsilon_t$, $\omega_t$ and  $\nu_t$ are white noise processes with mean zero and variances $\sigma^2_\epsilon$, $\sigma^2_\omega$ and $\sigma^2_\nu$, respectively.

%	The DLR is determined in state space form, with a prior distribution for the state vector at time $t=0$ and matrices expressed as follows:
%	
%\begin{align*}
%\boldsymbol{w_t'} = ( \nu_t, \omega_t, \epsilon_t, 0),~
%\boldsymbol{X_0} \sim \mathcal{N} (\boldsymbol{m_0}, \boldsymbol{C_0}), ~
%\boldsymbol{X_t} =
%\begin{bmatrix}
%\alpha_t & \beta_t
%\end{bmatrix},~
%\boldsymbol{f_t}= \begin{bmatrix}
%1 & \log(r_t)
%\end{bmatrix},~
%\boldsymbol{G_t} =
%\begin{bmatrix}
%	1 & 0 \\
%	0 & 1
%\end{bmatrix}
%\end{align*}
Similar models with more variables ca be used to forecast demand based on different variables, but, in our case, price plays a significant role and we can simplify the model to Equation (\ref{dlmobs}). It is notable that $r_t$ is not stochastic. The models was implemented in R using the \textit{`dlm'} package and and parameters are estimated with maximum likelihood \cite{petris2010r}.

%\subsection{Machine Learning models}

\subsection{Artificial Neural Network} \label{annmodel}

	An ANN is a supervised ML algorithm which is inspired by the human brain and learns from experience. An ANN consists of neurons and layers that are connected with arcs and input that is translated to output through an activation function. ANNs are powerful algorithms that are able to model any continuous, non-linear system, and can then make generalizations and predict the unseen values \cite{GuoqiangZhang1998}. ANNs are used extensively in various fields of forecasting such as demand forecasting in SC \cite{Aburto2007} and information sharing in SC \cite{Carbonneau2008}.

	In this paper, we employed the commonly used `feed-forward error back-propagation' type of ANN with one hidden layer \cite{Carbonneau2008}. In this type of ANN, the prices as inputs are entered to the neurons in each layer. Each neuron is connected to all of the neurons in the next layer. The aim is to minimize the error between the predicted and actual values of demand. It is vital to transfer the ANN data set before training the neural net. We transferred the data using the min-max method and considered price and demand time series as the inputs to forecast demand. A logistic function was used as the activation function. The model was fitted using the \textit{`neuralnet'} package in R \cite{gunther2010neuralnet}.

\subsection{Support Vector Regression}\label{svrmodel}
	SVR is a powerful supervised learning algorithm in which the output is a numerical variable. It is the most common application of support vector machines (SVM) \cite{Basak2007}. SVM implements intelligent algorithms to discover the patterns in complex data sets. SVR has been used in many different applications including demand forecasting  \cite{villegas2018support}.
	
	The difference between SVR and ordinary least square regression models is that SVR attempts to minimize the generalized error, whereas statistical regression models try to minimize the deviation of forecasted values from the actual ones. SVR finds a linear function in the space within a distance of $\epsilon$ from its predicted values. Any violation of this distance is penalized by a constant (C) \cite{Ali2009}. SVR employs a kernel function to transfer low dimensional data to high dimensional data. The kernel function can take on a number of different forms. We have used a non-linear function. We fitted the model using the `\textit{e1071}' package in R \cite{dimitriadou2006e1071}. Finally, cross-validation is done in order to choose the best combination of the kernel function, kernel coefficient and C. We ran $\gamma$ on a sequence of intervals of width 0.1 ranging from 0 to 1, and C on a sequence of intervals of width 1 ranging from 0 to 100. This generated a total of 1100 samples.

	\section{Data} \label{data}
%	The SC network consists of one manufacturer, a number of distributors and nearly two thousand retailers. Retailers scan the data at the point of sale (POS) and place their orders with the manufacturer.
 	We gathered data from a food manufacturing company producing hundreds of FMCG sin Australia. Consumers demand data are available from the point of sales and price are gathered by the company. Demand data are aggregated across the retailers and span 112 weeks. There is 843 stock keeping unit location (SKUL). Demand levels differ greatly between promotional and non-promotional periods. This difference is mainly due to the promotion impact and promotion contributes significantly to the CoV. Demand during promotions can be up to 60 times greater than those during non-promotional periods for highly volatile demand series. Depending on the demand size and promotional package, demand may be impacted over a number of periods before and after promotion. The natural logarithm has been employed to reduce the displayed variability of demand (throughout this paper, the natural logarithm of demand is used for modeling). Demands are heavily impacted with promotion and they are not of a seasonally demanded nature. 
%The forecasting practice in the industry is to apply triple exponential smoothing. Then, experts override the generated forecast judgmentally by taking into account promotional information. This is also a common approach in other industries \citep{wang2016select}.     

 	Since demand series are highly volatile, we have categorized them based on their CoV into three different groups. There are different categories to measure relative volatility of time series \cite{packowski2013lean,scholz2012integration}. There is no consensus on a cut-off value for CoV, rather it depends on the type of time series, related industry, and volume \cite{syntetos2005categorization}. We use \cite{scholz2012integration} classification approach to categorize the SKUL demands based on their CoV as follows:
\begin{itemize}
	\item Low volatility demand where CoV is smaller than 0.5. There are 311 SKULs in this category. The average CoV of these demands is 0.32.
	\item Moderate volatility where CoV is greater than 0.5 and smaller than one. There are 255 SKULs in this category. The average CoV of these demands series is 0.75.
	\item High volatility where CoV is greater than one. There are 277 SKULs in this category. The average CoV of these demands series is 1.71. 
\end{itemize}

Table \ref{analysis1} provides a summary of the descriptive statistics. As it can be seen in Table \ref{analysis1}, products have a large range of demands and are highly impacted by promotion. For example, while the average demand for the products with moderate volatility is 1763 units, the minimum demand is 31 units, and the maximum is 37,911. The 90\% quantile is 3273 units, meaning that 90\% of products have less than 3273 units of demand. Note that, this statistics is across all the SKULs in this category, but it can show the diversity of demand.

\begin{table}
	\centering	
	\caption{Descriptive statistics of sales}
	\label{analysis1}
	\vspace{0.5cm}
	\begin{tabular}{lcccccr}
		\hline
		& Mean &  Min & Max &Median& 75\% percentile&90\% percentile \\
		\hline
		Low volatility &1915&22 &25659&817 &1020 & 4314 \\
		
		Moderate volatility &1763 &31& 37911 & 752& 1397& 3273 \\
		
		High volatility & 1552&30&16162& 860& 3451& 3844\\
		\hline
	\end{tabular}
	
\end{table}

\begin{figure}[h] 
	\centering
	\includegraphics[scale=0.45] {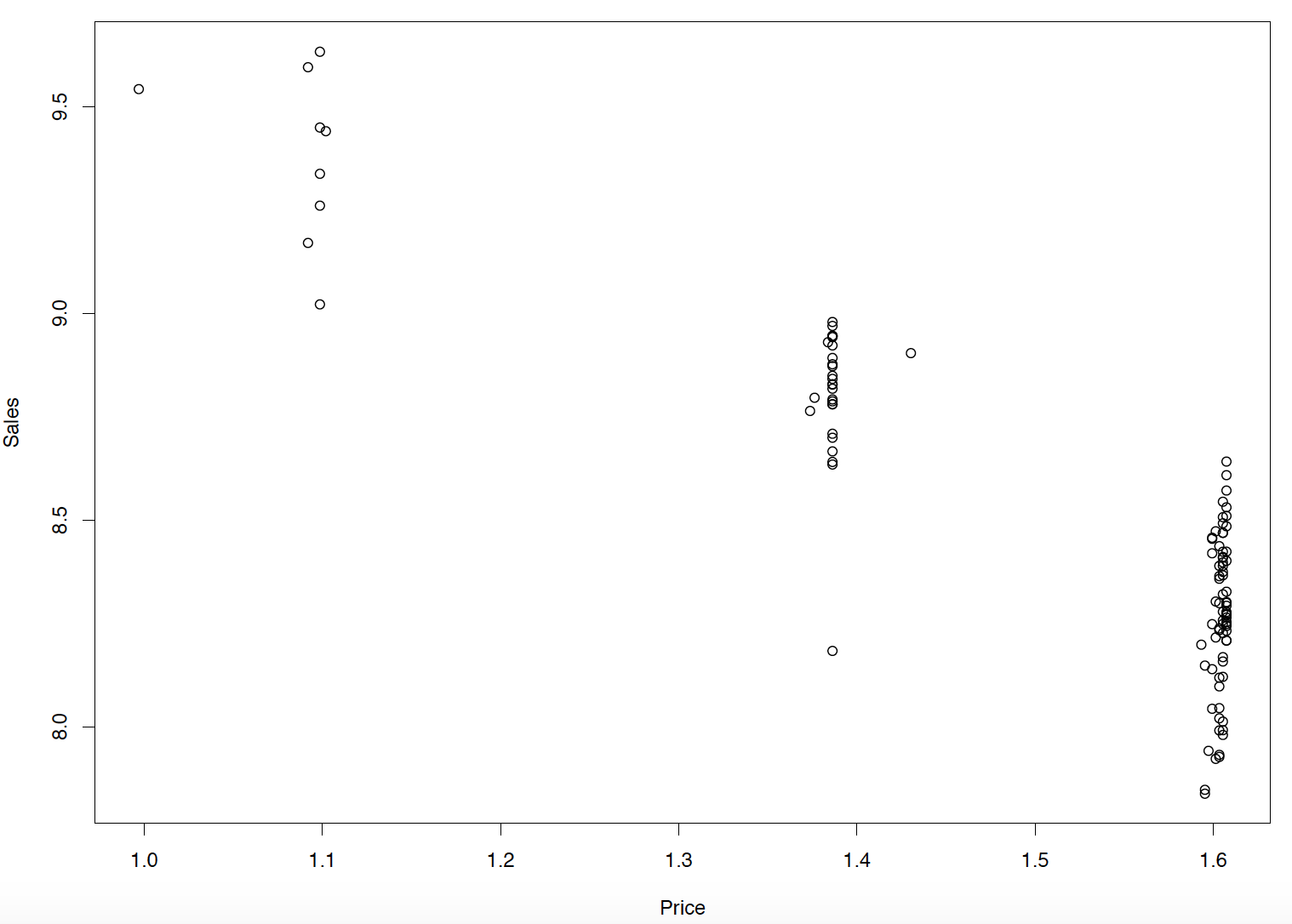}
	\caption{Sales-price relationship}
	\label{piece}
\end{figure}

Figure \ref{piece} shows the loglog sales-price relationship for a particular SKUL. Price is fixed over a number of prices, but sales vary over the time. Other SKULs have similar type of demand.

	\section{Empirical results and discussion}\label{resuts}
	This section presents the empirical results found using the methods presented in this paper. The data were split into a training-set and a test-set. The first 104 weeks were used as the training-set and estimate the model parameters. The last eight weeks are considered as the test-set and used to measure the performance and accuracy of the forecasting models. Since price is known in advance for the next eight weeks, we use it as an explanatory variable and generate forecast based on rolling origin for the next eight weeks. The ARIMA and ETS models are used as simple benchmarks. We also used the Theta model as another benchmark which is a hybrid model that benefits from decomposition and has been implemented successfully on M3 competition \cite{makridakis2000m3}. 
	
	Since the scale of the demand varies across series, we need to use scale-independent criteria to validate the accuracy of models  \cite{kolassa2007advantages}. Mean absolute scaled error (MASE) is used to evaluate the validity of each of the presented models \cite{Hyndman2006}. MASE is defined as follows:
	 
	 \begin{align} \label{MASE}
	 	MASE= \frac{1}{n} \frac{\sum_{t=1}^{n}|Y_t - \hat{Y}_t|}{\frac{1}{n-1}\sum_{t=2}^{n}|Y_t - Y_{t-1}|}
	 \end{align} 
	 
%	 \begin{align}\label{sMAPE}
%	 sMAPE= \frac{2}{n} \sum_{t=1}^{n} \frac{|Y_t -\hat{Y}_t|} {|Y_t + \hat{Y}_t|} * 100\% , 
%	 \end{align}
%	 where $n$ is the forecasting horizon, $Y_t$ is the actual value of the series, and $\hat{Y}_t$ is the generated forecast. 
	Tables \ref{1steptable} and \ref{1steptablepvalue} show the forecasting accuracy for the different models for eight-step ahead and p-values corresponding to each model performance across different values of CoV, respectively.
	
\begin{table}[h]
	\centering	
	\caption{Eight-step ahead forecasting accuracy}
%	\small\renewcommand{\arraystretch}{.9}
	\begin{tabular}{lcccr}
		\hline
		{Forecasting Models}&\multicolumn{3}{|c}{SKULs}\\
		\hline
		&CoV $\le$ 0.5 & 0.5 $<$ CoV $\le$1 & CoV $>$ 1 & Total\\
		\hline
		HR-ARIMA& 0.16 & 0.16&\textbf{0.17}& \textbf{0.18}\\ 
		\hline
		ARIMAX& \textbf{0.14}& \textbf{0.12} &0.39 & 0.21\\ 
		\hline
		ETSX &0.30& 0.51&1.40&0.69 \\ 
		\hline
		DLR   &0.19& 0.19 & 0.40& 0.25\\ 
		\hline
		SVR & 0.17 &0.19&0.41&0.25\\ 
		\hline
		ANN &0.19 &0.25 &1.10&1.00 \\ 
		\hline
		ARIMA &0.22 &0.30 &0.81&0.42\\ 
		\hline
		ETS &0.28 & 0.45& 1.07&0.57\\ 
		\hline
		Theta & 0.22 &0.30& 0.81&0.42\\ 
		\hline
			\end{tabular}
	\label{1steptable}
\end{table}

\begin{table} [H]
	\centering	
	\caption{P-values: Volatility impact on the forecasting accuracy(MASE) of models}
	\begin{tabular}{lcccr}
		\hline
		{Forecasting Models}&\multicolumn{4}{|c}{SKULs category based on CoV}\\
		\hline
		& &CoV $\le$ 0.5 & 0.5 $<$ CoV $\le$1 & CoV $>$ 1 \\
		\hline
		HR-ARIMA & CoV $\le$ 0.5& - & 0.16& 0.27\\ 
		& 0.5 $<$ CoV $\le$1& 0.16 & - &0.14\\ 
		& CoV $>$ 1 & 0.27 &  0.14&  -\\ 
		\hline
		ARIMAX & CoV $\le$ 0.5& - & 0.15& 0.05\\ 
		& 0.5 $<$ CoV $\le$1&0.15 & - &\textbf{0.01}\\ 
		& CoV $>$ 1 & 0.05 &  \textbf{0.01}&  -\\ 
		\hline
		ETSX & CoV $\le$ 0.5& - & \textbf{0.00}& \textbf{0.00}\\ 
		& 0.5 $<$ CoV $\le$1& \textbf{0.00}& - &\textbf{0.00}\\ 
		& CoV $>$ 1 &  \textbf{0.00}& \textbf{0.00} &  -\\ 
		\hline
		DLR  & CoV $\le$ 0.5& - & 0.18& \textbf{0.02}\\ 
		& 0.5 $<$ CoV $\le$1& 0.18& - &\textbf{0.00}\\ 
		& CoV $>$ 1 &  \textbf{0.02}& \textbf{0.00} &  -\\ 
		\hline
		
		SVR & CoV $\le$ 0.5& - &0.07 & 0.00\\ 
		& 0.5 $<$ CoV $\le$1&0.07 & - &\textbf{0.04}\\ 
		& CoV $>$ 1 &\textbf{0.00}  & \textbf{0.04} &  -\\ 
		\hline
		ANN & CoV $\le$ 0.5& - & \textbf{0.02}& \textbf{0.00}\\ 
		& 0.5 $<$ CoV $\le$1& \textbf{0.02}& - &\textbf{0.00}\\ 
		& CoV $>$ 1 & \textbf{0.00} &\textbf{0.00}  &  -\\ 
		\hline
		ARIMA & CoV $\le$ 0.5& - & \textbf{0.01}&\textbf{0.00} \\ 
		& 0.5 $<$ CoV $\le$1& \textbf{0.01}& - & \textbf{0.00}\\ 
		& CoV $>$ 1 & \textbf{0.00} & \textbf{0.00} &  -\\ 
		\hline
		ETS & CoV $\le$ 0.5& - &\textbf{0.00}&\textbf{0.00} \\ 
		& 0.5 $<$ CoV $\le$1& \textbf{0.00}& - &\textbf{0.00}\\ 
		& CoV $>$ 1 & \textbf{0.00}  & \textbf{0.00} &  -\\ 
		\hline
		Theta & CoV $\le$ 0.5& - &\textbf{0.01} & \textbf{0.00}\\ 
		& 0.5 $<$ CoV $\le$1& \textbf{0.01}& - &\textbf{0.00}\\ 
		& CoV $>$ 1 &  \textbf{0.00}& \textbf{0.00} &  -\\ 
		\hline
		\label{1steptablepvalue}
	\end{tabular}
\end{table} 

	The results for the eight-step ahead forecasts slightly differs across different categories and can be described as follows. For the series with low volatility, ARIMAX has the lowest MASE followed by HR-ARIMA. While adding covariates to ETS did not improve its accuracy,  it improved the accuracy of the ARIMA model significantly. All models except ETSX outperform the benchmarks. DLR and SVR showed robust performance and higher accuracy comparing to other statistical and ML models across different categories. While ANN works well for series with low volatility, its accuracy decreases dramatically when series have higher volatilities. These results show that this method requires further investigation since it contradicts other results found in the literature \cite{Ali2009,alon2001forecasting}. Defining new architecture for the ANN, identifying the input and most relevant type of data are some variables that may be considered to improve ANN results \cite{crone2010feature}.

	The results of Table \ref{1steptable} shows that HR-ARIMA model is the only robust model that performs well for different time series across all different volatilities. There is no significant difference in the accuracy of models when CoV is increasing. This shows that decomposition is helpful to deal with volatile time series. The models ARIMAX, DLR and SVR have shown robust performance between the low volatility and moderate volatility categories. However, there is a significant difference in their performance when CoV is high. The rest of the models are not robust when faced with volatile demand series. 

%The results for eight-week ahead forecasting differ from those obtained for one-step ahead forecasting. For the series with low volatility, ARIMAX and SVR outperformed all other models in terms of MASE. Also, the DLR has the smallest sMAPE.   Similarly, for the one-step ahead forecasts, adding a covariate to ETS did not exhibit superior performance. However, the added covariate to ARIMA model improved its performance. In long term forecasting, ETSX and ANN showed better performance in terms of sMAPE.

\begin{figure}[h] 
	\centering
	\includegraphics[scale=0.50] {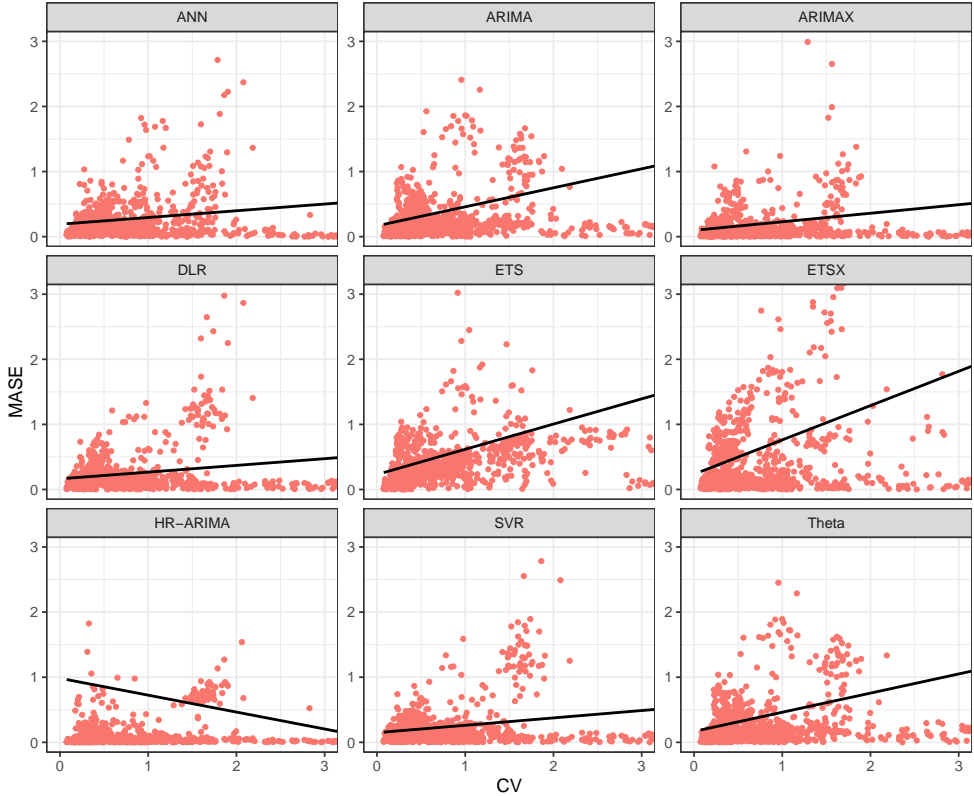}
	\caption{Models accuracy in eight-step ahead (MASE)- Each point represents one SKUL}
	\label{1stepMASEfinal}
\end{figure}

	Figures ~\ref{1stepMASEfinal} shows the scatter plot of eight-step ahead MASE with respect to their CoV for all SKULs for the different models studied. The magnitude of MASE varies for different models and increases as CoV increases. However, HR-ARIMA is more robust to CoV changes. This shows that decomposition is useful when the series show volatilities and hybrid models perform better for highly volatile demand series. Among the other models, DLR and SVR are more robust to CoV changes, whereas ETSX and ETS accuracy decrease when CoV is increasing. 

 	In general, of the statistical models considered in this section, the constructed hybrid regression time series model has the highest accuracy across moderate and highly volatile demands. This model uses Algorithm \ref{Heuristic Algorithm} and benefits from a piecewise regression and an ARIMA model to estimate promotional and non-promotional demands, respectively. The superior performance of the hybrid model is the result of decomposing sales into different components and using the appropriate model to forecast each of them. This is consistent with the other results found in the literature \cite{Pai2005}. Among statistical models, the ETSX model has a poor performance and is not recommended to use for forecasting volatile time series. On the other hand, DLR seems to be promising and SVR generates robust forecasts across different values of CoV. Our results are consistent with the results described by \cite{wang2009rule}. Surprisingly, the \textit{Theta} model did not generate robust and accurate forecasts across different categories and its accuracy decreases as CoV increases. This might be because we use an explanatory variable for other presented models in this thesis, however, there was no explanatory variable used by \textit{Theta} model.

	\section{Conclusion}\label{conc}

  Forecasting retail sales are of importance for many managerial decisions at different levels of supply chain. In the modern competitive market, many internal and external factors impact demand in different ways and make them volatile and unpredictable. Promotion is one of the factors that can have differing effects on demand dynamics and can make demand volatile not only over the promotional periods but also over the entire demand series.
 	
% 	depending on the promotion tools such as promotion type, promotion frequency, price cut-off and potential competitors in the market that are employed. There is a need for simple, yet accurate models to forecast promotional sales as sophisticated models are not commonly used in practice due to lack of expertise and resources \citep{hughes2001forecasting}. 
 	
 	We used CoV as an important, simple, and yet informative statistics to measure the relative volatility of demand time series that are impacted by promotion. We investigated the behavior of 843 SKULs that are impacted by promotions and have different levels of volatilities. We categorized demand series into three groups based on their CoV as low volatile, moderately volatile, and highly volatile. We showed that volatility of demands may significantly change the forecasting accuracy. Then, we empirically found the most suitable model for demand time series with different values of CoV.

	We decomposed demand time series into baseline demand and promotional demand (uplifts because of promotion). Then, we constructed a piecewise regression model which is able to model demand uplifts during promotional periods effectively, and an ARIMA model to forecast baseline demand. This model has a closed-form prediction interval and generates reliable forecasts when demand series are volatile. We also evaluated the different type of well-established models in the literature including ARIMA, ARIMAX, ETS, ETSX, DLR, \textit{Theta} and common ML algorithms, ANN and SVR to empirically evaluate their performance and robustness when demand series exhibit different levels of volatility.

%	Our empirical analysis shows that a hybrid model that captures the underlying behavior of time series by considering the driving factors of sales is the most appropriate approach to forecast highly volatile series. We proposed an ARIMA model to predict the baseline sales and a piecewise linear regression model that is appropriate to capture promotional sales uplifts. However, the proposed algorithm is generic and any other model can be developed for modeling promotional uplifts. 
	
	We showed that among the other presented models, ARIMAX has a superior performance for low and moderately volatile products. Surprisingly, adding covariate to ETS brings about no improvement in accuracy. DLR and SVR showed similar and robust performance across different values of CoV. ANN generates better results for low volatile demand series but works poor for highly volatile demand series. We also, showed that simple statistical models can outperform some of the sophisticated ML and statistical models. Our models can be used to forecast and optmize the required inventory level when demand series are highly volatile.
	
	For future research, there is a lot that needs both researchers and practitioners attention. As we have shown in this paper, different methods have different levels of performance under different levels of volatilities. One can analytically or empirically analyze why some models work better than others in certain conditions. An integrated approach which combines different techniques for promotional and non-promotional periods and encompasses all aspects of the relevant demand characteristics is a desirable direction for future research. The study focuses on a FMCG company in Australia's market, this may limit the implications of findings in different industries. The research models may be examined and validated in a different context with different data set. It would be beneficial to test different ML and heuristic forecasting methods which are able to estimate the model parameters precisely against the limited amount of available historical data for promotions. 
 	
 	There are many different variables that govern the dynamics of demand, but it could be expensive and complicated to consider all of them. Moreover, data may not be available to use. In this paper, the price is used as there is a strong correlation between price and demand and price explains the majority of variation in demand. Considering other influential factors such as other promotional information, special events, holidays and display types and using a multivariate model would be a worthwhile avenue of investigation. 
 	
 	 % 	There has been a few studies why some methods works better than others for certain type of  time series with certain characteristics \citep{kang2017visualising, wang2009rule}. 
 	 
% 	 Another interesting future research direction would be to find ways to determine the most appropriate forecasting method by considering multiple criteria. Some methods are more accurate for point forecasts, while others have less uncertainty and generate better prediction intervals. The determination of the best forecasting method by considering multiple criteria is an open problem in theory and practice.

% \section*{Acknowledgement(s)}
% 
% This study was funded by the Australian Research Council (Grant ID: IC140100032). The authors are grateful to Prof. Aris Syntetos and Prof. George Athanasopoulos for their invaluable feedback and constructive comments.
% 
% 	 
 	
 	% % % % % % % % % % % % % % % % % % % % % % % % % % % % % % % % % % % % % % % % %

%	\bibliographystyle{plainnat}
	\bibliographystyle{apalike}
	\bibliography{library2.bib}

\begin{thebibliography}{}

\bibitem[Aburto and Weber, 2007]{Aburto2007}
Aburto, L. and Weber, R. (2007).
\newblock Improved supply chain management based on hybrid demand forecasts.
\newblock {\em Applied Soft Computing}, 7(1):136--144.

\bibitem[Ahmed et~al., 2010]{ahmed2010empirical}
Ahmed, N.~K., Atiya, A.~F., Gayar, N.~E., and El-Shishiny, H. (2010).
\newblock An empirical comparison of machine learning models for time series
  forecasting.
\newblock {\em Econometric Reviews}, 29(5-6):594--621.

\bibitem[Ali et~al., 2009]{Ali2009}
Ali, {\"O}.~G., Say{\i}n, S., Van~Woensel, T., and Fransoo, J. (2009).
\newblock {SKU} demand forecasting in the presence of promotions.
\newblock {\em Expert Systems with Applications}, 36(10):12340--12348.

\bibitem[Alon et~al., 2001]{alon2001forecasting}
Alon, I., Qi, M., and Sadowski, R.~J. (2001).
\newblock Forecasting aggregate retail sales:: a comparison of artificial
  neural networks and traditional methods.
\newblock {\em Journal of Retailing and Consumer Services}, 8(3):147--156.

\bibitem[Armstrong, 2001]{armstrong2001principles}
Armstrong, J.~S. (2001).
\newblock {\em Principles of forecasting: a handbook for researchers and
  practitioners}, volume~30.
\newblock Springer Science \& Business Media.

\bibitem[Armstrong and Green, 2017]{armstrong2017demand}
Armstrong, J.~S. and Green, K.~C. (2017).
\newblock Demand forecasting {II}: Evidence-based methods and checklists.

\bibitem[Assimakopoulos and Nikolopoulos, 2000]{assimakopoulos2000theta}
Assimakopoulos, V. and Nikolopoulos, K. (2000).
\newblock The theta model: A decomposition approach to forecasting.
\newblock {\em International Journal of Forecasting}, 16(4):521--530.

\bibitem[Athanasopoulos and Hyndman, 2008]{athanasopoulos2008modelling}
Athanasopoulos, G. and Hyndman, R.~J. (2008).
\newblock Forecasting {Australian} domestic tourism.
\newblock {\em Tourism Management}, 29(1):19--31.

\bibitem[Aye et~al., 2015]{Aye2015}
Aye, G.~C., Balcilar, M., Gupta, R., and Majumdar, A. (2015).
\newblock Forecasting aggregate retail sales: The case of {South} {Africa}.
\newblock {\em International Journal of Production Economics}, 160:66--79.

\bibitem[Basak et~al., 2007]{Basak2007}
Basak, D., Pal, S., and Patranabis, D.~C. (2007).
\newblock Support vector regression.
\newblock {\em Neural Information Processing-Letters and Reviews},
  11(10):203--224.

\bibitem[Billah et~al., 2006]{Billah2006}
Billah, B., King, M.~L., Snyder, R.~D., and Koehler, A.~B. (2006).
\newblock Exponential smoothing model selection for forecasting.
\newblock {\em International journal of forecasting}, 22(2):239--247.

\bibitem[Blattberg and Neslin, 1993]{blattberg1993}
Blattberg, R.~C. and Neslin, S.~A. (1993).
\newblock Sales promotion models.
\newblock {\em Handbooks in Operations Research and Management Science},
  5:553--609.

\bibitem[Brown, 1959]{brown1959statistical}
Brown, R.~G. (1959).
\newblock {\em Statistical forecasting for inventory control}.
\newblock McGraw/Hill.

\bibitem[Cachon, 1999]{cachon1999managing}
Cachon, G.~P. (1999).
\newblock Managing supply chain demand variability with scheduled ordering
  policies.
\newblock {\em Management Science}, 45(6):843--856.

\bibitem[Caprio et~al., 1983]{caprio1983short}
Caprio, U.~D., Genesio, R., Pozzi, S., and Vicino, A. (1983).
\newblock Short term load forecasting in electric power systems: A comparison
  of {ARMA} models and extended {W}iener filtering.
\newblock {\em Journal of Forecasting}, 2(1):59--76.

\bibitem[Carbonneau et~al., 2008]{Carbonneau2008}
Carbonneau, R., Laframboise, K., and Vahidov, R. (2008).
\newblock Application of machine learning techniques for supply chain demand
  forecasting.
\newblock {\em European Journal of Operational Research}, 184(3):1140--1154.

\bibitem[Chan and Ripley, 2018]{TSA}
Chan, K.-S. and Ripley, B. (2018).
\newblock {\em TSA: Time Series Analysis}.
\newblock R package version 1.2.

\bibitem[Christen et~al., 1997]{Christen2016}
Christen, M., Gupta, S., Porter, J.~C., Staelin, R., and Wittink, D.~R. (1997).
\newblock Using market-level data to understand promotion effects in a
  nonlinear model.
\newblock {\em Journal of Marketing Research}, 34(3):322--334.

\bibitem[Christopher, 2000]{christopher2000agile}
Christopher, M. (2000).
\newblock The agile supply chain: competing in volatile markets.
\newblock {\em Industrial Marketing Management}, 29(1):37--44.

\bibitem[Christopher and Holweg, 2011]{christopher2011supply}
Christopher, M. and Holweg, M. (2011).
\newblock “supply chain 2.0”: Managing supply chains in the era of
  turbulence.
\newblock {\em International Journal of Physical Distribution \& Logistics
  Management}, 41(1):63--82.

\bibitem[Christopher and Holweg, 2017]{christopher2017supply}
Christopher, M. and Holweg, M. (2017).
\newblock Supply chain 2.0 revisited: a framework for managing
  volatility-induced risk in the supply chain.
\newblock {\em International Journal of Physical Distribution \& Logistics
  Management}, 47(1):2--17.

\bibitem[Collopy and Armstrong, 1992]{collopy1992rule}
Collopy, F. and Armstrong, J.~S. (1992).
\newblock Rule-based forecasting: Development and validation of an expert
  systems approach to combining time series extrapolations.
\newblock {\em Management Science}, 38(10):1394--1414.

\bibitem[Crone and Kourentzes, 2010]{crone2010feature}
Crone, S.~F. and Kourentzes, N. (2010).
\newblock Feature selection for time series prediction--a combined filter and
  wrapper approach for neural networks.
\newblock {\em Neurocomputing}, 73(10-12):1923--1936.

\bibitem[Dimitriadou et~al., 2006]{dimitriadou2006e1071}
Dimitriadou, E., Hornik, K., Leisch, F., Meyer, D., Weingessel, A., and Leisch,
  M.~F. (2006).
\newblock The e1071 package.
\newblock {\em Misc Functions of Department of Statistics (e1071), TU Wien}.

\bibitem[Divakar et~al., 2005]{Divakar2005}
Divakar, S., Ratchford, B.~T., and Shankar, V. (2005).
\newblock Practice prize article—chan4cast: A multichannel, multiregion sales
  forecasting model and decision support system for consumer packaged goods.
\newblock {\em Marketing Science}, 24(3):334--350.

\bibitem[Donohue, 2000]{donohue2000efficient}
Donohue, K.~L. (2000).
\newblock Efficient supply contracts for fashion goods with forecast updating
  and two production modes.
\newblock {\em Management Science}, 46(11):1397--1411.

\bibitem[Fildes et~al., 2008]{fildes2008forecasting}
Fildes, R., Nikolopoulos, K., Crone, S.~F., and Syntetos, A.~A. (2008).
\newblock Forecasting and operational research: a review.
\newblock {\em Journal of the Operational Research Society}, 59(9):1150--1172.

\bibitem[Fulcher et~al., 2013]{fulcher2013highly}
Fulcher, B.~D., Little, M.~A., and Jones, N.~S. (2013).
\newblock Highly comparative time-series analysis: The empirical structure of
  time series and their methods.
\newblock {\em Journal of the Royal Society Interface}, 10(83):20130048.

\bibitem[Ghiassi and Nangoy, 2009]{ghiassi2009dynamic}
Ghiassi, M. and Nangoy, S. (2009).
\newblock A dynamic artificial neural network model for forecasting nonlinear
  processes.
\newblock {\em Computers \& Industrial Engineering}, 57(1):287--297.

\bibitem[Gilliland, 2010]{gilliland2010business}
Gilliland, M. (2010).
\newblock {\em The business forecasting deal: exposing myths, eliminating bad
  practices, providing practical solutions}, volume~27.
\newblock John Wiley \& Sons.

\bibitem[G{\"u}nther and Fritsch, 2010]{gunther2010neuralnet}
G{\"u}nther, F. and Fritsch, S. (2010).
\newblock neuralnet: Training of neural networks.
\newblock {\em The R Journal}, 2(1):30--38.

\bibitem[Guo et~al., 2017]{guo2017double}
Guo, F., Diao, J., Zhao, Q., Wang, D., and Sun, Q. (2017).
\newblock A double-level combination approach for demand forecasting of
  repairable airplane spare parts based on turnover data.
\newblock {\em Computers \& Industrial Engineering}, 110:92--108.

\bibitem[Hope and Fraser, 2003]{hope2003beyond}
Hope, J. and Fraser, R. (2003).
\newblock {\em Beyond budgeting: how managers can break free from the annual
  performance trap}.
\newblock Harvard Business Press.

\bibitem[Huang et~al., 2008]{huang2008demand}
Huang, M.-G., Chang, P.-L., and Chou, Y.-C. (2008).
\newblock Demand forecasting and smoothing capacity planning for products with
  high random demand volatility.
\newblock {\em International Journal of Production Research},
  46(12):3223--3239.

\bibitem[Hyndman, 2018]{hyndman2007automatic}
Hyndman, R. (2018).
\newblock {\em fpp2: Data for "Forecasting: Principles and Practice" (2nd
  Edition)}.
\newblock R package version 2.3.

\bibitem[Hyndman et~al., 2008]{hyndman2008forecasting}
Hyndman, R., Koehler, A.~B., Ord, J.~K., and Snyder, R.~D. (2008).
\newblock {\em Forecasting with exponential smoothing: the state space
  approach}.
\newblock Springer Science \& Business Media.

\bibitem[Hyndman, 2006]{Hyndman2006}
Hyndman, R.~J. (2006).
\newblock Another look at forecast-accuracy metrics for intermittent demand.
\newblock {\em Foresight: The International Journal of Applied Forecasting},
  4(4):43--46.

\bibitem[Hyndman and Athanasopoulos, 2014]{hyndman2014forecasting}
Hyndman, R.~J. and Athanasopoulos, G. (2014).
\newblock {\em Forecasting: principles and practice}.
\newblock OTexts.

\bibitem[Jetta and Rengifo, 2011]{Jetta2011}
Jetta, K. and Rengifo, E.~W. (2011).
\newblock A model to improve the estimation of baseline retail sales.
\newblock {\em Journal of Centrum Cathedra}, 4(1):10--26.

\bibitem[Jordan and Messner, 2019]{jordan2019use}
Jordan, S. and Messner, M. (2019).
\newblock The use of forecast accuracy indicators to improve planning quality:
  Insights from a case study.
\newblock {\em European Accounting Review}, pages 1--23.

\bibitem[Jung and Jeong, 2012]{jung2012managing}
Jung, H. and Jeong, S.-J. (2012).
\newblock Managing demand uncertainty through fuzzy inference in supply chain
  planning.
\newblock {\em International Journal of Production Research},
  50(19):5415--5429.

\bibitem[Kolassa et~al., 2007]{kolassa2007advantages}
Kolassa, S., Sch{\"u}tz, W., et~al. (2007).
\newblock Advantages of the mad/mean ratio over the mape.
\newblock {\em Foresight: The International Journal of Applied Forecasting},
  (6):40--43.

\bibitem[Lawrence et~al., 2006]{lawrence2006judgmental}
Lawrence, M., Goodwin, P., O'Connor, M., and {\"O}nkal, D. (2006).
\newblock Judgmental forecasting: A review of progress over the last 25 years.
\newblock {\em International Journal of forecasting}, 22(3):493--518.

\bibitem[Makridakis and Hibon, 2000]{makridakis2000m3}
Makridakis, S. and Hibon, M. (2000).
\newblock The {M3}-competition: results, conclusions and implications.
\newblock {\em International Journal of Forecasting}, 16(4):451--476.

\bibitem[Makridakis et~al., 2018]{makridakis2018}
Makridakis, S., Spiliotis, E., and Assimakopoulos, V. (2018).
\newblock Statistical and machine learning forecasting methods: Concerns and
  ways forward.
\newblock {\em PLOS One}, 13(3):e0194889.

\bibitem[Meade, 2000]{meade2000evidence}
Meade, N. (2000).
\newblock Evidence for the selection of forecasting methods.
\newblock {\em Journal of Forecasting}, 19(6):515--535.

\bibitem[Meindl and Chopra, 2001]{chopra2004supply}
Meindl, P. and Chopra, S. (2001).
\newblock {\em Supply chain management: Strategy, planning, and operation}.
\newblock Prentice Hall.

\bibitem[Min, 2010]{min2010artificial}
Min, H. (2010).
\newblock Artificial intelligence in supply chain management: Theory and
  applications.
\newblock {\em International Journal of Logistics: Research and Applications},
  13(1):13--39.

\bibitem[Munaron, 2017]{volatilitythesis}
Munaron, V. (2017).
\newblock {\em Analysis of Demand Volatility}.
\newblock {Master of Science} dissertation, Delft University of Technology.

\bibitem[Narayanan et~al., ]{narayanandemand}
Narayanan, A., Sahin, F., and Robinson, E.~P.
\newblock Demand and order-fulfillment planning: The impact of point-of-sale
  data, retailer orders and distribution center orders on forecast accuracy.
\newblock {\em Journal of Operations Management}.

\bibitem[Nikolopoulos et~al., 2015]{Nikolopoulos2015}
Nikolopoulos, K., Litsa, A., Petropoulos, F., Bougioukos, V., and Khammash, M.
  (2015).
\newblock Relative performance of methods for forecasting special events.
\newblock {\em Journal of Business Research}, 68(8):1785--1791.

\bibitem[Packowski, 2013]{packowski2013lean}
Packowski, J. (2013).
\newblock {\em LEAN Supply Chain Planning: The New Supply Chain Management
  Paradigm for Process Industries to Master Today's {VUCA} World}.
\newblock CRC Press.

\bibitem[Pai and Lin, 2005]{Pai2005}
Pai, P.-F. and Lin, C.-S. (2005).
\newblock A hybrid {ARIMA} and support vector machines model in stock price
  forecasting.
\newblock {\em Omega}, 33(6):497--505.

\bibitem[Petris and An, 2010]{petris2010r}
Petris, G. and An, R. (2010).
\newblock An {R} package for dynamic linear models.
\newblock {\em Journal of Statistical Software}, 36(12):1--16.

\bibitem[Petris et~al., 2009]{dlmr}
Petris, G., Petrone, S., and Campagnoli, P. (2009).
\newblock Dynamic linear models.
\newblock {\em Springer}.

\bibitem[Priore et~al., 2018]{priore2018applying}
Priore, P., Ponte, B., Rosillo, R., and de~la Fuente, D. (2018).
\newblock Applying machine learning to the dynamic selection of replenishment
  policies in fast-changing supply chain environments.
\newblock {\em International Journal of Production Research},
  57(11):3663--3677.

\bibitem[Ramanathan, 2012]{Ramanathan2012}
Ramanathan, U. (2012).
\newblock Supply chain collaboration for improved forecast accuracy of
  promotional sales.
\newblock {\em International Journal of Operations \& Production Management},
  32(6):676--695.

\bibitem[Ramanathan and Muyldermans, 2011]{Ramanathan2011}
Ramanathan, U. and Muyldermans, L. (2011).
\newblock Identifying the underlying structure of demand during promotions: A
  structural equation modelling approach.
\newblock {\em Expert Systems With Applications}, 38(5):5544--5552.

\bibitem[{Robert C . Blattberg} and J, 1995]{Science2016a}
{Robert C . Blattberg}, R.~B. and J, E. (1995).
\newblock {How Promotions Work}.
\newblock {\em Marketing Science}, 14(3):122--132.

\bibitem[Scholz-Reiter et~al., 2012]{scholz2012integration}
Scholz-Reiter, B., Heger, J., Meinecke, C., and Bergmann, J. (2012).
\newblock Integration of demand forecasts in abc-xyz analysis: Practical
  investigation at an industrial company.
\newblock {\em International Journal of Productivity and Performance
  Management}, 61(4):445--451.

\bibitem[Serdarasan, 2013]{serdarasan2013review}
Serdarasan, S. (2013).
\newblock A review of supply chain complexity drivers.
\newblock {\em Computers \& Industrial Engineering}, 66(3):533--540.

\bibitem[Shah, 1997]{shah1997model}
Shah, C. (1997).
\newblock Model selection in univariate time series forecasting using
  discriminant analysis.
\newblock {\em International Journal of Forecasting}, 13(4):489--500.

\bibitem[Sillanp{\"a}{\"a} and Liesi{\"o}, 2018]{sillanpaa2018forecasting}
Sillanp{\"a}{\"a}, V. and Liesi{\"o}, J. (2018).
\newblock Forecasting replenishment orders in retail: value of modelling low
  and intermittent consumer demand with distributions.
\newblock {\em International Journal of Production Research},
  56(12):4168--4185.

\bibitem[Silver et~al., 1998]{silver1998inventory}
Silver, E.~A., Pyke, D.~F., Peterson, R., et~al. (1998).
\newblock {\em Inventory management and production planning and scheduling},
  volume~3.
\newblock Wiley New York.

\bibitem[Svetunkov, 2018]{ivan}
Svetunkov, I. (2018).
\newblock smooth: Forecasting using state space models.
\newblock R package version 2.4.5.

\bibitem[Syntetos et~al., 2016]{syntetos2016supply}
Syntetos, A.~A., Babai, Z., Boylan, J.~E., Kolassa, S., and Nikolopoulos, K.
  (2016).
\newblock Supply chain forecasting: Theory, practice, their gap and the future.
\newblock {\em European Journal of Operational Research}, 252(1):1--26.

\bibitem[Syntetos et~al., 2005]{syntetos2005categorization}
Syntetos, A.~A., Boylan, J.~E., and Croston, J. (2005).
\newblock On the categorization of demand patterns.
\newblock {\em Journal of the Operational Research Society}, 56(5):495--503.

\bibitem[Torkul et~al., 2016]{torkul2016real}
Torkul, O., Y{\i}lmaz, R., Selvi, I., and Cesur, M.~R. (2016).
\newblock A real-time inventory model to manage variance of demand for
  decreasing inventory holding cost.
\newblock {\em Computers \& Industrial Engineering}, 102:435--439.

\bibitem[Trapero et~al., 2015]{Trapero2014}
Trapero, J.~R., Kourentzes, N., and Fildes, R. (2015).
\newblock On the identification of sales forecasting models in the presence of
  promotions.
\newblock {\em Journal of the operational Research Society}, 66(2):299--307.

\bibitem[Van~Heerde et~al., 2002]{van2002promotions}
Van~Heerde, H.~J., Leeflang, P.~S., and Wittink, D.~R. (2002).
\newblock How promotions work: {SCAN}* {PRO}-based evolutionary model building.
\newblock {\em Schmalenbach Business Review}, 54(3):198--220.

\bibitem[Villegas et~al., 2018]{villegas2018support}
Villegas, M.~A., Pedregal, D.~J., and Trapero, J.~R. (2018).
\newblock A support vector machine for model selection in demand forecasting
  applications.
\newblock {\em Computers \& Industrial Engineering}, 121:1--7.

\bibitem[Walker and Weber, 1984]{walker1984transaction}
Walker, G. and Weber, D. (1984).
\newblock A transaction cost approach to make-or-buy decisions.
\newblock {\em Administrative Science Quarterly}, pages 373--391.

\bibitem[Wang et~al., 2006]{wang2006characteristic}
Wang, X., Smith, K., and Hyndman, R. (2006).
\newblock Characteristic-based clustering for time series data.
\newblock {\em Data Mining and Knowledge Discovery}, 13(3):335--364.

\bibitem[Wang et~al., 2009]{wang2009rule}
Wang, X., Smith-Miles, K., and Hyndman, R. (2009).
\newblock Rule induction for forecasting method selection: Meta-learning the
  characteristics of univariate time series.
\newblock {\em Neurocomputing}, 72(10-12):2581--2594.

\bibitem[Zhang et~al., 1998]{GuoqiangZhang1998}
Zhang, G., Patuwo, B.~E., and Hu, M.~Y. (1998).
\newblock Forecasting with artificial neural networks:: The state of the art.
\newblock {\em International Journal of Forecasting}, 14(1):35--62.

\end{thebibliography}

\newpage

\end{document}